\documentclass[3p,12pt]{elsarticle}
\usepackage{graphicx}
\usepackage{amsmath}
\usepackage{amssymb}

\begin{document}

\newcommand{\be}{\begin{equation}}
\newcommand{\ee}{\end{equation}}

\newcommand{\rd}{\mathrm{d}}
\newcommand{\rme}{\mathrm{e}}
\newcommand{\ri}{\mathrm{i}}

\renewcommand{\th}{\tilde{h}}
\newcommand{\tr}{\tilde{r}}
\newcommand{\tx}{\tilde{x}}
\newcommand{\txi}{\tilde{\xi}}
\newcommand{\tR}{\tilde{R}}
\newcommand{\tRp}{\tilde{R}'}
\newcommand{\tQ}{\tilde{Q}}
\newcommand{\teta}{\tilde{\eta}}
\newcommand{\bep}{\bar\varepsilon}
\newcommand{\tl}{\tilde{\Lambda}}

\begin{frontmatter}

\title{Static and dynamic response of a fluid-fluid interface to electric point and line charge}

\author[ntnu]{Simen {\AA}. Ellingsen\corref{cor}\fnref{fax}}
\ead{simen.a.ellingsen@ntnu.no}
\author[ntnu]{Iver Brevik}
\ead{iver.h.brevik@ntnu.no}

\cortext[cor]{Corresponding author.}
\fntext[fax]{Fax number: +47 73593491. Telephone: +47 73593554.}
\address[ntnu]{Department of Energy and Process Engineering, Norwegian University of Science and Technology, N-7491 Trondheim, Norway}

\bigskip
Revised version, June 2012

\begin{abstract}
  We consider the behavior of a dielectric fluid-fluid interface in the presence of a strong electric field from a point charge and line charge, respectively, both statically and, in the latter case, dynamically. The fluid surface is elevated above its undisturbed level until balance is reached between the electromagnetic lifting force, gravity and surface tension. We derive ordinary differential equations for the shape of the fluid-fluid interface which are solved numerically with standard means, demonstrating how the elevation depends on field strength and surface tension coefficient. In the dynamic case of a moving line charge, the surface of an inviscid liquid-liquid interface is left to oscillate behind the moving charge after it has been lifted against the force of gravity. We show how the wavelength of the oscillations depends on the relative strength of the forces of gravity and inertia, whereas the amplitude of the oscillations is a nontrivial function of the velocity at which the line charge moves.
\end{abstract}

\end{frontmatter}

\section{Introduction}
Electromagnetic manipulation of dielectric soft matter surfaces has emerged to be a topic of considerable interest in later years, both from fundamental and applied points of view. The integration of photonics and microfluidics, \emph{optofluidics}, is a new analytic field - cf., for instance, the  reviews \cite{fan11,monat07,psaltis06}.

One of the first examples of this kind was the classic experiment of Ashkin and Dziedzic \cite{ashkin73} on the elevation of a water surface when illuminated vertically from above by a pulsed laser.
The force per unit area acting on a dielectric boundary in the presence of an electric field is \cite{brevik79} ${ P}_E=\int {\bf f}_E \cdot d{\bf n}$, where
\be\label{force}
  {\bf f}_E=-\frac{\varepsilon_0}{2}E^2\nabla \varepsilon
\ee
is the force density in the boundary layer.
The surface elevation of the surface was found to be small, of the order of 1 $\mu$m, due to the strong surface tension between air and water \cite{ashkin73}. Another noteworthy example on the interaction between light and matter is the experiment of Zhang and Chang \cite{zhang88}, showing how the surface of a micrometer water droplet when illuminated by a laser
becomes distorted, especially at the rear end because of the lens effect of the sphere.
A newer series of experiments is provided by Delville {\it et al.} \cite{casner04,delville09,wunenburger06,wunenburger11}, working with a two-fluid system of surfactant-coated nanodroplets in oil microemulsions near the critical point. The surface tension $\sigma$ can then be reduced to less than a millionth of the usual air-water tension, and the surface displacements can accordingly be very large, almost 100 $\mu$m. The classical electromagnetic theory of this sort of experiments is given in several papers; cf., e.g., \cite{hallanger05} and the review article \cite{delville06}.

The above remarks are related to optical manipulation of liquid surfaces. A close sibling of optofluidics is to make use of {\it electrostatic fields} instead of optical fields. This field of research, often termed electrohydrodynamics, is not new --- cf., for instance, Taylor's seminal 1969 paper \cite{taylor69} --- yet in view of the considerable attention given to the optical techniques it is still of interest to understand the electrostatic case. Naturally, the electric force density acting in a boundary layer is basically the same as above, viz.\ Eq.~(\ref{force}). Since an optical period is much shorter than the mechanical response time of a fluid, optofluidic theory works with time-averaged fields (over an optical period), and similarities are many. But there are obvious differences also. For one thing, the permittivity of a dielectric fluid is usually strongly frequency dependent, and is often much higher in the static case than in the high-frequency optical case. The typical mode of approach in electrohydrodynamics has involved a rather heavy, primarily numerical, framework \cite{castellanos98} where keeping track of the various physical effects can be challenging. We analyse here a simple realizable case in which quite rich physics can be described by relatively simple means, allowing a treatment which is physically transparent throughout.

Formation of liquid columns and extraction of jets by electric fields were studied in seminal experiments of Raco \cite{raco68} and Taylor \cite{taylor69}, and have been a source of investigation at intervals since then. Electrically supported liquid columns have been subjected to stability investigations both experimentally and theoretically\cite{saville70,sankaran92,burcham00,pelekasis01}, and the formation of such columns rising against the gravitational force have been shown experimentally \cite{terasawa83,dong01}.

The experiment of Raco \cite{raco68}, in particular, is extreme in that it is very nonlinear and moreover is influenced by the finite conductivity of the liquid. In order to describe such an experiment, a heavy computational machinery is likely to be necessary. However, a useful insight into the fluid-field interaction may be obtained also by analytical methods, on the basis of quite simple models. Our intention is just to describe the fluid interface response with respect to a strong electric field by taking an external point charge $Q$, or a charged line with charge $\Lambda$ per unit length,  to be the source. This simple model has a high degree of flexibility and its analysis is physically transparent as well as mathematically tractable. To some extent, such a theory is related to that given by de Gennes \cite{degennes05} on the pushing of a micro droplet using a laser beam.

We will assume in the following that the typical height $h$ of the surface elevation is much smaller than the distance $L$ between the charge and the undisturbed surface. The reason is that, while the equation giving the fluid deformation in a given electric field is simple to derive, the complementary problem of how the electric field from a point charge is altered by the presence of an arbitrarily shaped dielectric surface is a more difficult problem best solved within computational physics. As is so often the case, however, the problem is analytically tractable when restricted to leading order in the small quantity $h/L$, which allows calculation of the field at an arbitrary point by method of image charges as if the surface were at rest. Inspection of the governing equations which we derive in the following sections is enough to see that including corrections to the electric fields due to the surface perturbations provides a higher order correction. For consistency we must also linearize the mechanical equations of motion for the surface. The problem is nevertheless nontrivial and physically interesting, and illuminates key features in the theory of electric liquid surface manipulation.

Note  that it is safe to ignore non-linear dielectric response due to the quadratic electro-optic effect in the following. C.f.\ e.g.\ discussion on p.~147 of Ref.~\cite{brevik79}.

Finally, to put our work into perspective, let us compare it with some earlier works. The present area of research is  obviously interdisciplinary, encompassing electromagnetism, optics, fluid mechanics, and chemistry. It is fair to say that most earlier studies have been concerned with the optical approach - in addition to the references given above, we mention here also the extensive paper of Issenmann {\it et al.}\cite{issenmann11} on unsteady deformations of a free liquid surface caused by radiation pressure. Sakai and Yamamoto \cite{sakai06}  developed an interesting field tweezers method for characterization of a liquid surface; here  a needle electrode was used to make  the deformation of the surface. Another noncontact method was invented by Sohi {\it et al.}  \cite{sohi78} on how electrocapillarity can be used to generate surface waves that in turn are detected via specular reflection of a laser beam from the fluid surface. A further development, now on the chemistry side, is the study by Stenvot and Langevin \cite{stenvot88} of surface viscoelasticity of monolayers by means of an analysis of propagating capillary waves.

The chief characteristic of the present paper is that we use an electric {\it point} or a {\it line charge} to displace the surface, and then calculate the surface elevation using the Maxwell stress tensor. The advantage of the model is that it is concrete, and reasonably simple to handle mathematically.  In the case of a line charge we also analyse the situation where the line moves with constant velocity parallel to the surface, although in the last case we restrict ourselves to an inviscid fluid and neglect surface tension. In subsection \ref{sec_st} we discuss semi-quantitatively the more general situation in which surface tension plays a role in the formation of gravity-capillary waves. The latter case is quite important physically - it shows interesting analogies to electrodynamic phenomena such as the \v{C}erenkov effect - and would deserve a separate study of its own. This lies outside the scope of the present paper, however.

\section{Surface elevation arising from a point charge}\label{sec:pointcharge}

Consider the setup sketched in Fig.~1: a point charge $Q$ is situated a distance $L$ above the undisturbed liquid surface (the $xy$ plane). In cylindrical coordinates, the position of $Q$ is $r=0, z=L$. The permittivity in the upper region  is $\varepsilon_1$, and in the lower region  $\varepsilon_2$. We assume that  $\varepsilon_2 > \varepsilon_1$ so that the surface force  acts upwards. Now introduce the image charge
\begin{equation}
Q'=-\frac{\varepsilon_2-\varepsilon_1}{\varepsilon_2+\varepsilon_1}Q, \label{2}
\end{equation}
situated at $r=0, z=-L$. The potential $\phi_1(r, z)$ at an arbitrary field point P in the upper region  is that of the charge $Q$ plus the fictitious charge $Q'$ in a uniform medium of permittivity $ \varepsilon_1$ \cite{landau84},
\begin{equation}
\phi_1=\frac{1}{4\pi \varepsilon_0}\left(\frac{Q}{\varepsilon_1 R}+\frac{Q'}{\varepsilon_1 R'}\right), \label{3}
\end{equation}
where $R$ and $R'$ are the distances QP and Q'P,
\begin{equation}
R=\sqrt{r^2+(L-z)^2}, \quad R'=\sqrt{r^2+(L+z)^2}. \label{4}
\end{equation}
Furthermore, if $Q''$ is the image
\begin{equation}
Q''=\frac{2\varepsilon_2}{\varepsilon_2+\varepsilon_1}Q, \label{5}
\end{equation}
situated at the same position as $Q$, the potential $\phi_2$ in the lower region is that of $Q''$ in a uniform medium of permittivity $\varepsilon_2$,
\begin{equation}
\phi_2=\frac{1}{4\pi \varepsilon_0}\frac{Q''}{\varepsilon_2 R}. \label{6}
\end{equation}
These expressions satisfy the boundary conditions at $z=0$, namely that  $\phi$ as well as of $\varepsilon \partial \phi/\partial z$ are continuous.

\begin{figure}[htb]
  \centering\includegraphics[width=3in]{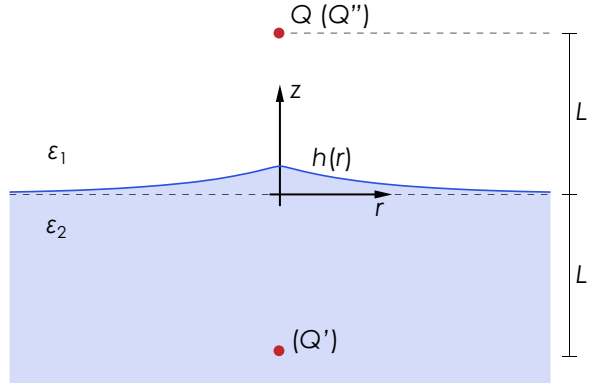}
  \caption{The geometry considered in section \ref{sec:pointcharge}.}
  \label{fig:pointgeometry}
\end{figure}

The radial component $E_{1r}=-\partial \phi_1/\partial r$ and the vertical component $E_{1z}=-\partial \phi_1/\partial z$ of the field in the upper region are
\begin{eqnarray}
E_{1r}&=&\frac{1}{4\pi \varepsilon_0\varepsilon_1}\left( \frac{Qr}{R^3}+\frac{Q'r}{R'^3}\right), \label{7}\\
E_{1z}&=&\frac{1}{4\pi \varepsilon_0\varepsilon_1}\left( -Q\frac{L-z}{R^3}+Q'\frac{L+z}{R'^3}\right), \label{8}
\end{eqnarray}
which gives for the square $E_1^2=E_{1r}^2+E_{1z}^2$ at an arbitrary position $r,z$ the expression
\begin{equation}
E_1^2=\frac{1}{(4\pi \varepsilon_0\varepsilon_1)^2}\left[ \frac{Q^2}{R^4}+\frac{Q'^2}{R'^4}-\frac{2QQ'}{R^3R'^3}(L^2-r^2-z^2)\right]. \label{9}
\end{equation}
Consider now the elevated surface, $h=h(r)$. The electric pressure acting on it is \cite{brevik79}
\begin{eqnarray}
{\bf P}_E&=&\frac{1}{2}\varepsilon_0(\varepsilon_2-\varepsilon_1)\left[ \frac{\varepsilon_2}{\varepsilon_2}E_{1n}^2+E_{1t}^2\right]\bf n \nonumber\\
&=&\frac{1}{2}\varepsilon_0(\varepsilon_2-\varepsilon_1)\left[\frac{\varepsilon_1}{\varepsilon_2}E_1^2+\left(1-\frac{\varepsilon_1}{\varepsilon_2}\right)E_{1t}^2\right]\bf n, \label{10}
\end{eqnarray}
where the normal field ${\bf E}_{1n}$ and the tangential field ${\bf E}_{1t}$ are evaluated on the outer surface, $z=h^+$, $\bf n$ being the outward normal vector
\be
  \mathbf{n}= \frac{-h'\hat{\mathbf{r}} + \hat{\mathbf{z}}}{\sqrt{1+h^{\prime 2}}}; ~~ \mathbf{t}= \frac{\hat{\mathbf{r}} + h'\hat{\mathbf{z}}}{\sqrt{1+h^{\prime 2}}}.
\ee
Here $\mathbf{t}$ is a tangential unit vector and $h'=h'(r)$ is the slope of the surface. Thus
\begin{equation}
E_{1t}=\frac{1}{\sqrt{1+h'^2}}(E_{1r}+E_{1z}h'), \label{11}
\end{equation}
which is easily calculable from Eqs.~(\ref{7}) and (\ref{8}) inserting $z=h$.

For computational purposes it is convenient to introduce here a reference field strength $E_0$, equal to the field due  to $Q$ only, at the origin $r=0, z=0$,
\begin{equation}
E_0=\frac{1}{4\pi \varepsilon_0\varepsilon_1}\frac{Q}{L^2}. \label{12}
\end{equation}
We can then write, at position $z=h^+$,
\begin{eqnarray}
  E_1^2&=&E_0^2\left[ \frac{1}{(R/L)^4}+\frac{(Q'/Q)^2}{(R'/L)^4}+\frac{2Q'/Q}{(R/L)^3(R'/L)^3}\left( \frac{r^2+h^2}{L^2}-1\right)\right], \label{13}\\
  E_{1t}^2&=&\frac{E_0^2}{1+h'^2}\left[ \frac{ (r/L)-(1-h/L)h'}{(R/L)^3}+\frac{Q'}{Q}\,\frac{(r/L)+(1+h/L)h'}{(R'/L)^3}\right]^2 \label{14}
\end{eqnarray}
(observe that $E_{1n}^2=E_1^2-E_{1t}^2$).

For numerical purposes we use the dimensionless quantities
\be\label{dimless}
  \th=\frac{h}{L},~~\tr = \frac{r}{L},~~ \tR=\frac{R}{L};~~\tRp=\frac{R'}{L};~~ \tQ = \frac{Q'}{Q}=-\frac{\varepsilon_2-\varepsilon_1}{\varepsilon_2+\varepsilon_1}.
\ee
Noting $\th' = \rd \th/\rd \tr = h'$ we may write to leading order in $h,h'$
\begin{eqnarray}
  E_1^2&=& \frac{E_0^2}{(\tr^2+1)^2}\left(1+\tQ^2+2\tQ\frac{\tr^2-1}{\tr^2+1}\right)+\frac{4(1-\tQ)}{(\tr^2+1)^3}\th+...\\
  E_{1t}^2&=&\frac{E_0^2}{(\tr^2+1)^3}\left[\tr^2(1+\tQ)^2+2\tr(1-\tQ^2)\left(\frac{3\tr}{\tr^2+1}\th-\th'\right) \right]+...
\end{eqnarray}
plus corrections of order $\th^2, \th\th'$.

The excess pressure force due to surface-tension is related to the two radii of curvature $R_1$ and $R_2$ at a given point by\cite{landau87}
\begin{equation}
  {\bf P}_\mathrm{s.t.}=p_{\rm s.t.}\mathbf{n}=\sigma\left(\frac1{R_1}+\frac1{R_2}\right)\bf n. \label{16}
\end{equation}
We use the sign convention of \cite{landau87} that the curvature is positive when the surface is concave upwards, which implies that near the maximum of the elevation, the radii are negative, and the surface tension force points downwards. With a bit of differential geometry this may be written in terms of $h$ and its derivatives as \cite{wehausen60}
\begin{equation}
  p_{\rm s.t.}=\frac{\sigma}{r}\,\frac{d}{dr}\frac{rh'}{\sqrt{1+h'^2}} = \frac{\sigma}{r}(h'+rh'')+... \label{16}
\end{equation}
where $\sigma$ is the surface-pressure coefficient. The force of gravity is accounted for by noting that a surface elevated by a distance $h$ has a hydrostatic pressure which is lower than that at the unperturbed surface by a difference
\be
  \Delta p_\mathrm{hydrostat.} = - (\rho_2-\rho_1)gh
\ee
giving a negative upward (i.e., downward) surface force. Now note that all surface forces are directed normally to the surface. The balance equation is therefore scalar and the stationary situation is found when the net surface force density is zero. Altogether, the governing equation for the surface elevation takes the form
\begin{equation}\label{pointEqDim}
  (\rho_2-\rho_1)gh-p_\mathrm{s.t.}=\frac{1}{2}\varepsilon_0(\varepsilon_2-\varepsilon_1)\left[ \frac{\varepsilon_1}{\varepsilon_2}E_1^2+\left(1-\frac{\varepsilon_1}{\varepsilon_2}\right) E_{1t}^2\right].
\end{equation}
As the density $\rho_2$ of the lower liquid is larger than the density $\rho_1$ of the upper one, both terms on the left hand side of this equation are positive near $\tr=0$, thus both adding to the hydrostatic pressure at the level $z=0$. This is counterbalanced by the field pressure on the right hand side, acting upwards.

The boundary conditions to equation (\ref{pointEq}) are
\be\label{pointBC}
  \th(\infty)=\th'(0)=0.
\ee

\subsection{Numerical and large-$r$ asymptotic solutions}

Explicitly, thus, the dimensionless surface elevation $\th$ is given as the solution to the second order differential equation
\be\label{pointEq}
  \th''+[\tr^{-1}+Af_1(\tr)]\th'+ [Af_0(\tr)-G]\th + As(\tr)=0
\ee
where
\begin{eqnarray}
  f_1(\tr)&=& - \frac{2\tr (1-\bar\varepsilon)(1-\tQ^2)}{(\tr^2+1)^3},\\
  f_0(\tr)&=& \frac{6\tr^2(1-\bar\varepsilon)(1-\tQ^2)}{(\tr^2+1)^4}+\frac{4\bar\varepsilon A(1-\tQ)}{(\tr^2+1)^3},\\
  s(\tr)&=& \frac{(1+\tQ)^2\tr^2 + \bep (1-\tQ)^2}{(\tr^2+1)^3}.
\end{eqnarray}
and dimensionless parameters
\be\label{dimlessAG}
  A = \frac{\varepsilon_0(\varepsilon_2-\varepsilon_1)E_0^2L}{2\sigma};~~ G=\frac{L^2}{l_c^2}; ~~ \bar\varepsilon=\frac{\varepsilon_1}{\varepsilon_2}.
\ee
As is typical for problems involving deformation of liquid surfaces, we may introduce the capillary length
\be
  l_c = \sqrt{\frac{\sigma}{(\rho_2-\rho_1)g}}.
\ee
The quantity $A$ is a relative measure of the influence of the electromagnetic force compared to that from surface tension. $G$, similarly, is a measure of the force of gravity relative to the surface tension force. In macroscopic geometries surface tension can often play a very minor role, so both $A$ and $G$ may take values much greater than unity.
For a water-air interface, $l_c\approx 2.8$mm \cite{landau87}.

\begin{figure}[tb]
  \centering\includegraphics[width=3in]{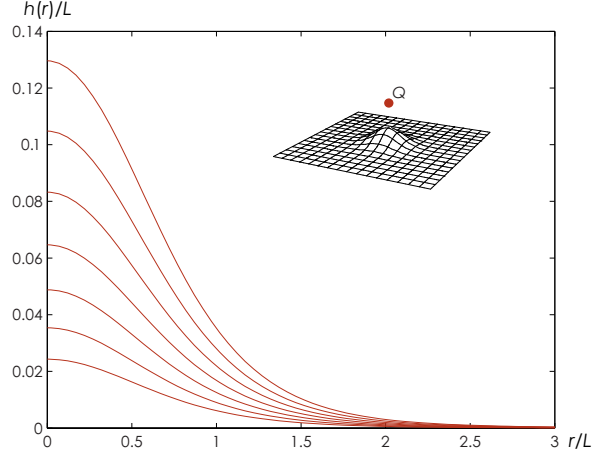}
  \caption{Surface elevation for different point charge field strengths (bottom to top) $E_0=5\cdot 10^5$V/m to $1.1\cdot 10^6$V/m in steps of $10^5$V/m. This means $Q$ ranges from $5.6$ to $12.2$ nC. Parameters are $L=1$cm, $\varepsilon_1=1, \varepsilon_2=15, (\rho_2-\rho_1)g=9790$N/m$^3$. The inset figure is to illustrate the geometry and is not to scale.}
  \label{fig:pointchargeelevation}
\end{figure}

We now turn to the behaviour of $h(r)$ for large $r$. Letting $\tr\to \infty$ we have
\be
  \frac1{\tr}+Af_1\sim \frac1{\tr},~~ Af_0-G\sim -G, ~~ As\sim \frac{A(1+\tQ)^2}{\tr^4},
\ee
so if we assume $\th$ falls off as a power of $\tr$, $\th\propto \tr^{-\mu}$, then $f_1\th'\propto \th'' \propto \tr^{-\mu-2}\ll f_0\th$, so these terms can be neglected in Eq.~(\ref{pointEq}). The asymptotic solution is then trivially found as
\be\label{pointasymp}
  \th(\tr)\sim \frac{A(1+\tQ)^2}{G\tr^4} = \frac{\varepsilon_0\varepsilon_1E_0^2}{g(\rho_2-\rho_1)L}\frac{\varepsilon_2-\varepsilon_1}{\varepsilon_2+\varepsilon_1}\frac1{\tr^4},~~ \tr\to \infty.
\ee
The asymptotic solution is independent of $\sigma$ as one would expect in an area where the elevation varies slowly with $\tr$. It is determined by the ratio $A/G$ which is the relative influence of electromagnetic forces and gravity, which is also physically obvious. For numerical purposes it is more useful to use this asymptotic expansion as a boundary condition than $h(\infty)=0$, since the latter requires needlessly large values of $\tr$ be used.

The numerical solution of equation (\ref{pointEq}) with boundary conditions (\ref{pointBC}) is readily found, and shown in figure \ref{fig:pointchargeelevation}. We use straightforward Runge-Kutta integration with a shooting loop to find the value of $h(0)$ so that the asymptotic value (\ref{pointasymp}) is met at some large value of $r/L$.

\section{Surface elevation arising from a line charge}\label{sec:linecharge}

 Consider now the analogous situation where an infinite charged straight line lies parallel to the undisturbed surface (the $xy$ plane).  The line is parallel to the $y$ axis, and intersects the $zx$ plane in the point $x=0, z=L$.  We let now $\Lambda$ denote the charge per unit length. The images $\Lambda'$ and $\Lambda''$ are related to $\Lambda$ exactly as $Q'$ and $Q''$ were to $Q$ in the previous section, by expressions (\ref{2}) and (\ref{5}). The potentials are \cite{landau84}
 \begin{equation}
 \phi_1=-\frac{\Lambda}{2\pi \varepsilon_0\varepsilon_1}\ln R -\frac{\Lambda'}{2\pi \varepsilon_0\varepsilon_1}\ln R', \label{18}
 \end{equation}
 \begin{equation}
 \phi_2=-\frac{\Lambda'}{2\pi \varepsilon_0\varepsilon_2}\ln R, \label{19}
 \end{equation}
 where now $R$ and $R'$ are the distances in the $xz$ plane,
 \begin{equation}
 R=\sqrt{x^2+(L-z)^2}, \quad R'=\sqrt{x^2+(L+z)^2}. \label{20}
 \end{equation}
 We let henceforth $\eta=\eta(x)$ denote the surface elevation. The electric pressure ${\bf P}_E$ is given by the same formula (\ref{10}) as before, where now ${\bf E}_{1t}$ is the tangential component of ${\bf E}_1$ at the upper surface in the $zx$ plane. An analogous calculation to the one above gives for the desired field quantities, at $z=\eta^+$,
\begin{eqnarray}
  E_1^2&=&E_0^2\left[ \frac{1}{(R/L)^2}+\frac{\Lambda'^2}{\Lambda^2}\,\frac{1}{(R'/L)^2}+\frac{2\Lambda'/\Lambda}{(R/L)^2(R'/L)^2}
  \left(\frac{x^2+\eta^2}{L^2}-1\right)\right], \label{21}\\
  E_{1t}^2&=&\frac{E_0^2}{(1+\eta'^2)}\left[\frac{(x/L)-(1-\eta/L)\eta'}{(R/L)^2}+\frac{\Lambda'}{\Lambda}\,\frac{(x/L)
  +(1+\eta/L)\eta'}{(R'/L)^2}\right]^2, \label{22}
\end{eqnarray}
where $\eta' \equiv d\eta/dx$ and we have introduced the reference field strength
\begin{equation}
E_0=\frac{1}{2\pi \varepsilon_0\varepsilon_1}\,\frac{\Lambda}{L}, \label{23}
\end{equation}
in analogy to Eq.~(\ref{12}).

The surface-tension pressure is now \cite{landau87}
\begin{equation}
{\bf P}_{\rm s.t.}=p_{\rm s.t.}\mathbf{n}=\sigma \frac{\eta''}{(1+\eta'^2)^{3/2}}\,\mathbf{n} \sim (\sigma \eta'' +...)\mathbf{n},
\end{equation}
thus a simpler expression than the previous expression (\ref{16}). The governing equation thus becomes very similar to the one we found for the point charge
\begin{equation}\label{lineEq}
  (\rho_2-\rho_1)g\eta-p_{\rm s.t.} = \frac{1}{2}\varepsilon_0(\varepsilon_2-\varepsilon_1)\left[ \frac{\varepsilon_1}{\varepsilon_2}E_1^2+\left(1-\frac{\varepsilon_1}{\varepsilon_2}\right) E_{1t}^2\right],
\end{equation}
with Eqs.~(\ref{21}) and (\ref{22}) to be inserted on the right hand side. Boundary conditions are as before,
\be
  \teta(\infty)=\teta'(0)=0.
\ee

\begin{figure}[tb]
  \centering\includegraphics[width=3in]{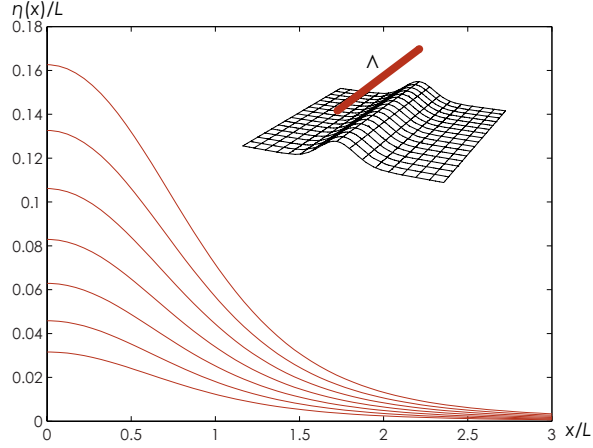}
  \caption{Surface elevation for different line charge field strengths for the same parameters as in Fig.~\ref{fig:pointchargeelevation} and the same values of $E_0$. This corresponds to $\lambda$ now ranging from $5.6$ to $12.2$ nC/cm.}
  \label{fig:linechargeelevation}
\end{figure}

\subsection{Numerical and asymptotic solution}
For numerical purposes we again introduce the non-dimensional quantities (\ref{dimless}) and (\ref{dimlessAG}). After linearising equation (\ref{lineEq}) it now reads
\be
  \teta'' +Af_1(x)\teta' + [Af_0(x)-G]\teta + As(x)=0
\ee
with $\teta=\eta/L$, $\tx=x/L$ and
\begin{eqnarray}
  f_1(\tx)&=& -\frac{2(1-\bep)(1-\tl^2)\tx}{(\tx^2+1)^2};\label{f1line}\\
  f_0(\tx)&=& \frac{2(1-\tl^2)(2\tx^2+\bep-\bep\tx^2)}{(\tx^2+1)^3};\\
  s(\tx)&=& \frac{(1+\tl)^2\tx^2+\bep(1-\tl)^2}{(\tx^2+1)^2}. \label{sline}
\end{eqnarray}
Here $\tl=\Lambda'/\Lambda=\tQ$.

For large $\tx$ we proceed as before. We have as $x\to \infty$
\be
  As\sim \frac{A(1+\tl)^2}{\tx^2};~~Af_0-G\sim -G; ~~Af_1\sim \frac{2A(1-\bep)(1-\tl^2)}{\tx^3}
\ee
so, assuming $\teta\propto \tx^{-\mu}$, $f_1\teta'\ll \teta''\ll \teta$, we again have a simple asymptotic expression
\be\label{lineAsymp}
  \teta \sim \frac{A(1+\tl)^2}{G\tx^2}=\frac{\varepsilon_0\varepsilon_1E_0^2}{g(\rho_2-\rho_1)L}\frac{\varepsilon_2-\varepsilon_1}
  {\varepsilon_2+\varepsilon_1}\frac1{\tx^2},~~ \txi\to \infty.
\ee
Example numerical calculations are shown in figure \ref{fig:linechargeelevation} for different electric field strengths, calculated in exactly the same way as for the point charge case.

\subsection{The role of surface tension}

\begin{figure}[tb]
  \centering\includegraphics[width=3in]{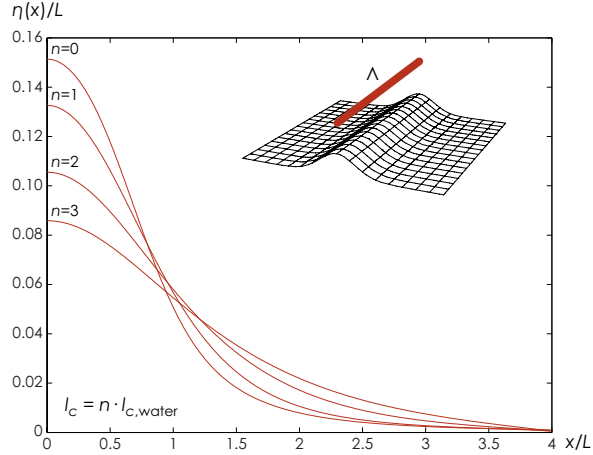}
  \caption{Surface elevation for different values of surface tension. Together with the zero surface tension case we plot the results of assuming once twice and thrice the capillary length of water, $l_{c,\mathrm{water}}=2.8$mm. Parameters are otherwise as in figure \ref{fig:linechargeelevation} with $E_0=10^6$V/m.}
  \label{fig:surfacetension}
\end{figure}

Before we move on to the dynamical case, let us briefly discuss the role of surface tension. The effect is well illustrated by figure \ref{fig:surfacetension} where we plot the elevation with a line charge with four different values of the capillary length: once, twice and thrice that of water as well as no surface tension. The surface tension force seeks to smoothen the surface, making the local radius of curvature as large as possible. Near the maximum of the elevation curve, surface tension seeks to pull the surface down, whereas the direction of the surface tension force is opposite in the regions where the curve is concave up. In general, surface tension counteract any tendency of a surface to curve. This is the reason why only small deformations were observed in experiment \cite{ashkin73} where surface tension was important, as opposed to the laser experiments of Delville's group \cite{casner04,delville09,wunenburger11} who are able to decrease surface tension by many orders of magnitude by operating close to a critical point and thereby effectuate large and strongly nonlinear interface deformations.

\section{Dynamical response in the two-dimensional case}

The dynamical response of a liquid-liquid interface to a time-varying radiation pressure is of obvious physical interest. There are several papers dealing with this topic, theoretical and experimental, when the laser illumination is fixed in space. In addition to the references given in Section 1, the recent, extensive treatment of Wunenburger {\it et al.} \cite{wunenburger06} is a key reference in that context.

A natural extension of the kind of static theory laid out in the previous sections is to consider the case where the exterior source is not fixed in space, but is moving with a constant velocity $V$ in the $x$ direction, parallel to the interface (the $xy$ plane) at a vertical separation $L$. As mentioned above, such a situation has already been considered by de Gennes \cite{degennes05} for practical, droplet-pushing purposes. As it is easiest to envisage the situation both conceptually and mathematically in the two-dimensional case, we will assume that  a charged line, directed in the $y$ direction,  is the moving source. That is, we adopt the same physical model as in Section \ref{sec:linecharge}.
Our purpose is to examine the time-dependent elevation of the interface, for a given value of $V$. We will here assume zero viscosity in the liquids, which means that there is no wake formation behind the line moving through medium 1 (the line is taken to have zero width). We assume $V$ is non-relativistic, and that the magnetic field from the moving charge is thus negligible compared to the electric.

In the discussions in this chapter we shall omit the effects of both viscosity and surface tension in order to make the physical picture more transparent and also simplify the mathematics somewhat. Inclusion of both these physical corrections is straightforward using the same analytical techniques as presented herein, however, and could form a useful extension of it.

Before we commence, we make the physically natural \emph{ansatz} that all physical quantities depend on $x$ and $t$ only through the Galilei transformation variable $\xi=x-Vt$. This is the same as assuming that in a coordinate system where the charge is at rest, the surface elevation appears stationary. With this we may replace in the following
\be\label{ansatz}
  \frac{\partial }{\partial x} = \frac{\rd}{\rd \xi};~~~\frac{\partial }{\partial t} = -V\frac{\rd}{\rd \xi}.
\ee
This ansatz is that used implicitly by Rapha\"{e}l and de Gennes \cite{raphael96}, but they keep their analysis more general by allowing an explicit time dependence. The ansatz made here allows us to find quasistationary solutions, but precludes any information about how such solutions may be accessed physically, i.e.\ how the system is turned on. On the other hand our solution has the advantage of not considering Fourier modes, which turns out to yield information about the quasistationary wave pattern quite readily.

\begin{figure}[tb]
  \centering\includegraphics[width=2.1in]{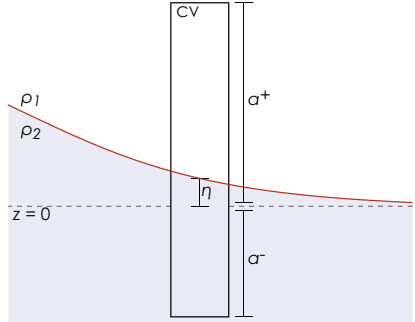}
  \caption{The geometry of the small control volume used to derive the equation of motion in the dynamic case. }
  \label{fig:CV}
\end{figure}

Consider now a tall and thin control volume CV as shown in figure \ref{fig:CV}, intersecting the interface with a small end area $\mathcal A$ whose extension is so small that $\eta(\xi)$ is approximately linear inside CV. We let $\eta$ mean the elevation in the center of CV. Newton's second law for such a volume, which follows from Reynolds' transport theorem, asserts that the $z$ component of the sum of external forces (here pressure and gravity) equals the change of fluid momentum within CV (e.g., Chapter 3 of \cite{white03}):
\be\label{reynolds}
  \sum F^\mathrm{ext}_{z}=\oint_\mathrm{\partial CV}\rho w \mathbf{v}\cdot\rd \mathbf{A} + \frac{\partial}{\partial t}\int_\mathrm{CV} \rho w \rd \mathcal V
\ee
Here we let the decomposition of the velocity vector be $\mathbf{v}=u\hat{\mathbf{x}}+w\hat{\mathbf{z}}$, $\rd\mathbf{A}$ is directed out of the control surface $\partial$CV, and $\rd\mathcal V$ denotes volume integration.

The sum of the external forces, including gravity, hydrostatic pressure and the pressure discontinuity at the surface due to electromagnetic forces and surface tension, is easily ascertained to be given by the same expressions as used in the previous sections,
\be\label{Fext}
    \sum F^\mathrm{ext}_{z} = \mathcal A[p_\mathrm{EM}+p_\mathrm{s.t.}-(\rho_2-\rho_1) g \eta]
\ee
and we need not derive this again. $p_\mathrm{EM}$ and $p_\mathrm{s.t.}$ have the same meaning as in section \ref{sec:linecharge}, but with $\xi$ replacing $x$ in $p_\mathrm{EM}$.

Treating the right hand side of equation (\ref{reynolds}) is a little more involved. First, let the heights $a^-$ and $a^+$ (see figure \ref{fig:CV}) be large, so that $w$ is zero at the end surfaces (clearly, $w$ must tend to zero far from the interface). To linear order there is no flux across the side surfaces of $\partial$CV. That disposes of the surface integral. We are left with
\be\label{zint}
  \sum F^\mathrm{ext}_{z}(\xi) = \mathcal A\rho_2\frac{\partial}{\partial t}\int_{-\infty}^\eta w(\xi,z)\rd z + \mathcal A\rho_1\frac{\partial}{\partial t}\int_\eta^\infty w(\xi,z)\rd z.
\ee
We now require knowledge of how the $z$-velocity $w$ depends on $x$ and $z$. Since the surface elevation is assumed small, we are free to make use of the velocity potential $\phi$, where $\mathbf{v}=\nabla\phi$. Assuming incompressibility gives the continuity condition $\nabla^2\phi=0$ which upon further differentiation w.r.t.\ $z$ implies
\be\label{replacement}
  \Bigl(\frac{\partial^2}{\partial x^2}+\frac{\partial^2}{\partial z^2}\Bigr)w=0.
\ee
As boundary conditions we clearly must demand that $w(\xi,z=\pm\infty)=0$. Furthermore the so-called kinematic boundary condition at a surface is
\be\label{kinematicBC}
 w(\xi,z=\eta)=\dot\eta + u \frac{\rd \eta}{\rd x} \approx \dot \eta,
\ee
since the second term in the middle form is of quadratic order in $\eta/L$.

According to the standard method of separation of variables we assume $w(\xi,z)=\Xi(\xi)Z(z)$ and conclude that $\Xi''=-k^2 \Xi$ and $Z''=k^2 Z$ for some real number $k$. In Fourier form the solution may then be written
\be\label{Fourier}
  w(\xi,z) = VL\int_0^\infty \frac{\rd k}{2\pi}w(k)\rme^{-k|z-\eta|}\rme^{\ri k\xi}
\ee
which is seen to solve (\ref{replacement}) and fulfil boundary conditions at infinite $|z|$. Of course, the physical value is the real part of this. We next define the complex elevation
\be
  \eta_c(\xi) = L^2\int_0^\infty \frac{\rd k}{2\pi}\eta(k)\rme^{\ri k\xi}
\ee
so the kinematic boundary condition (\ref{kinematicBC}) gives $w(k)=-\ri k L\eta(k)$. With the form (\ref{Fourier}) the integrals in equation (\ref{zint}) are easy to solve, and inserting the answer into the left hand side of (\ref{Fext}) after differentiating using the ansatz (\ref{ansatz}) we obtain the equation
\begin{eqnarray}
  p_\mathrm{EM}(\xi,\eta_c,\eta_c')+p_\mathrm{s.t.}(\eta_c'')-(\rho_2-\rho_1) g \eta_c &=& -  V^2 L^2(\rho_1+\rho_2)\int_0^\infty \frac{\rd k}{2\pi} k\eta(k)\rme^{\ri k\xi}\nonumber \\  &=&
  \ri V^2 (\rho_1+\rho_2)\eta_c'.\label{newEq}
\end{eqnarray}

Now we may let $\eta_c=\eta + \ri \eta_i$ and write down the real and imaginary parts of equation (\ref{newEq}) separately. To simplify the picture somewhat, consider the system when there is no surface tension, $\sigma=0$. For the parameters used in the previous sections, this hardly changes the results, but makes the physics more transparent and also simplifies the numerical treatment considerably since stiffness is avoided and the order of the differential problem is reduced from 4 to 2. 
On the other hand, possible solutions of the equation may then be lost -- see section \label{sec_st} for a discussion of this point. 
Eliminating $\eta_i$ from the equations we obtain our governing equation without viscosity or surface tension,
\be
  \Bigl(f_1\frac{\rd}{\rd \xi}+f_0+\Gamma\Bigr)\Bigl(f_1\teta'+f_0\teta+s-\Gamma\teta\Bigr)=-\Upsilon^2 \teta''.
\ee
In order to have nondimensional parameters not containing $\sigma$, we have defined here
\be
  \Gamma = \frac{G}{A}=\frac{2gL(\rho_2-\rho_1)}{\varepsilon_0(\varepsilon_2-\varepsilon_1)E_0^2}; ~~ \Upsilon = \frac{2V^2(\rho_2+\rho_1)}{\varepsilon_0(\varepsilon_2-\varepsilon_1)E_0^2}.
\ee
Obviously these parameters are measures of the relative strength of gravitational to electromagnetic forces, and inertial to electromagnetic forces, respectively.
Explicitly,
\begin{eqnarray}\label{bigEq}
  (f_1^2+\Upsilon^2)\teta''+(f_1f_1'+2f_1f_0 - 2\Gamma f_1)\teta' 
  + [f_1f_0'+(f_0-\Gamma)^2]\teta + f_1s'+(f_0-\Gamma)s=0.
\end{eqnarray}
where $f_1(\txi),f_0(\txi),s(\txi)$ (argument $\txi$ suppressed above) are as defined in (\ref{f1line})-(\ref{sline}) upon $x\to \xi$. We have nondimensionalized as before $\txi=\xi/L$.

\subsection{Asymptotics and numerical solution}

Repeating our previous exercise of taking the large $\xi$ asymptotics of the different coefficient functions we find that the asymptotic form of $\teta(\xi)$ is unchanged from the static case if we assume the solution to fall off as a power law as $\xi^{-\mu}$ at infinity,
\be\label{lineAsymp}
  \teta(\xi) \sim \frac{s(\xi)}{\Gamma}\sim\frac{\varepsilon_0\varepsilon_1E_0^2}{g(\rho_2-\rho_1)L}\frac{\varepsilon_2-\varepsilon_1}{\varepsilon_2+\varepsilon_1}\frac1{\txi^2},~~ \txi\to \infty.
\ee
The asymptotic form of $\teta'$ is thus $\teta'\sim -2\teta/\txi^3$, which together form a useful set of boundary conditions at large $\xi$. This condition must clearly hold \emph{in front} of the moving charge, in accordance with our assumption that the liquid surface is at rest before the arrival of the line charge (with nonzero $\sigma$ included, forward-propagating waves are possible).

However, since our liquid is assumed inviscid, $\teta$ no longer needs to tend to zero at infinity now that we consider a system which is not in mechanical equilibrium. To wit, behind the moving line charge we might expect waves to form since the charge, after setting the surface in motion, moves on and leaves it to oscillate in its wake. When not assuming $\eta$ to tend to zero, the asymptotic equation from (\ref{bigEq}) must also include a $\teta''$ term,
\be\label{wave}
  \Upsilon^2\teta'' + \Gamma^2\teta \sim \Gamma \frac{(1+\tl)^2}{\txi^2}.
\ee
We shall not dwell on the particular solution to this equation (which can be found!), but note rather that equation (\ref{wave}) has two homogeneous solution which do not vanish at infinity,
\be
  \teta = C_1 \cos \frac{\Gamma \txi}{\Upsilon} + C_2 \sin \frac{\Gamma \txi}{\Upsilon}
\ee
with integration constants $C_1$ and $C_2$ which we are not in a position to evaluate in a simple way. This asymptotic form must be valid far \emph{behind} the moving charge where the surface has been disturbed and moves as an undamped wave.

\begin{figure}[tb]
  \centering\includegraphics[width=4.5in]{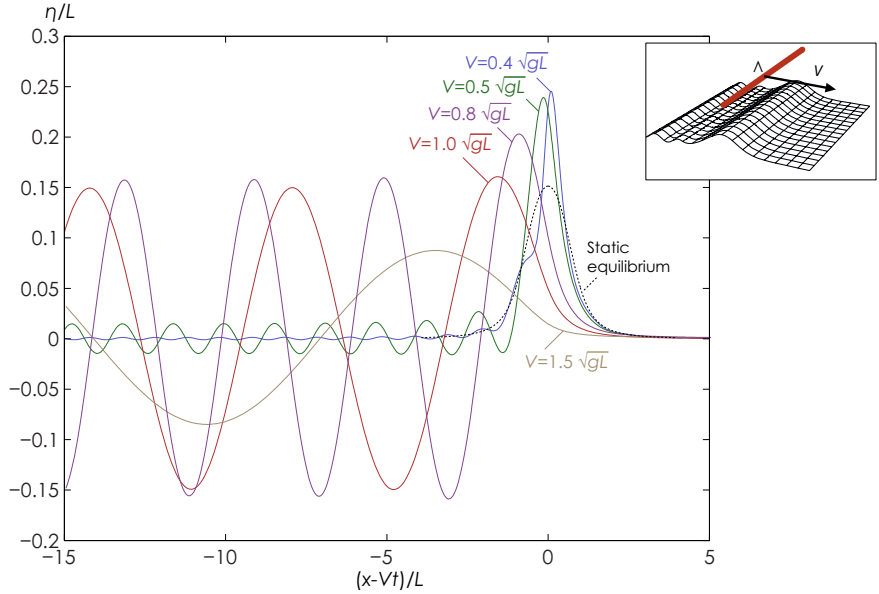}
  \caption{Surface elevation in the absence of surface tension and viscosity for a charge moving at velocity $V$ towards the right [particle position $(x,z)=(Vt,L)$]. We let $\rho_2 g=9790$N/m$^3$ and $\rho_1\approx 0$ as in the case of a water/air interface. The dotted line shows the static mechanical equilibrium situation (also  for $\sigma=0$).}
  \label{fig:dynamic}
\end{figure}

Notice at this point how our theory is valid when $V=0$ (the static case considered before) and when $V$ is sufficiently large, but \emph{not} when $V/\sqrt{gL}$ is small. The reason is that we have assumed not only that $\teta\ll 1$, but also $\teta'\ll 1$, in order to get rid of the terms $(1+\teta'^2)^{-1/2}$. However, when $V$ is small, the wavelength of the waves in the wake becomes short,
\be\label{lambda}
  \lambda = \frac{2\pi\Upsilon}{\Gamma}L=2\pi \frac{V^2}{g}\frac{\rho_2+\rho_1}{\rho_2-\rho_1}
\ee
and the derivative becomes proportional to $gLV^{-2}$. Note that this wavelength is independent of the electromagnetic properties, and is simply a measure of the relative strength of the forces of inertia and gravity.

In figure \ref{fig:dynamic} we plot the solution for a line charge moving towards the right at different velocities. At large $V$ there are no problems, but already at the two lowest values of $V$ we are in fact pushing the boundaries of linear theory, since $\teta'$ goes as high as $0.5$. Equation (\ref{bigEq}) predicts very high maximum elevation for small velocities, but this is wrong and happens because the compensating factor, such as the square root denominator in equation (\ref{22}), is not included. The highest peaks may thus be slightly overestimated in figure \ref{fig:dynamic}.

We observe a number of interesting features. For example, when velocity is small, the peak is in fact higher than the static, mechanical equilibrium value. This is because the rising liquid continues beyond the equilibrium value before the influence of gravity finally makes it turn and fall back again. On the other hand, when the line charge moves fast, the fluid only reaches its maximum velocity some time after the charge has moved on. We also see that the \emph{amplitude} of the harmonic oscillations is a non-trivial function of $V$. When the movement is very slow, oscillations behind the main peak are very small, since the charge, after lifting the surface, stays around to set it gently back down again. This is what we see in figure \ref{fig:dynamic} for the two slowest velocities. When $V$ is \emph{very} high, likewise, oscillations must, in addition to having long wavelength, be of small amplitude, since the surface hardly has time to feel the presence of the charge before it is gone again. Somewhere in between these extremes is found the velocity which gives the maximum oscillation amplitude. Finally, we see that for small velocities, the first peak appears \emph{in front} of the line charge, as is the case for $V=0.4\sqrt{gL}$ in figure \ref{fig:dynamic}. For even smaller velocities than this, the surface is lifted up, then drops again due to gravity, whereupon it is lifted again to a second time, since the field from the charge is still present and strongly felt. The beginnings of a second peak are observed in the $0.4\sqrt{gL}$ graph (to fully study this phenomenon, nonlinear theory must be invoked as previously discussed). All in all, our simple set-up treated with linear theory has yielded surprisingly rich physics.

In the previous we have omitted the effects of viscosity, although it is obvious what the effect of its inclusion would be. We have seen that far behind the moving charge, the governing equation takes the form of an undamped harmonic oscillator (or wave) equation. A viscosity term, being proportional to $\dot \eta$ would include a damping in this equation, making sure the harmonic waves be damped exponentially behind the moving charge, admitting overdamped and underdamped solutions depending on the relative value of viscosity and wave number $\Gamma/\Upsilon$. Studying this effect is straightforward within the above framework [a friction force is simply added to the right hand side of equation (\ref{Fext})]. Along with the effects of surface tension it serves as a useful extension problem for pedagogical purposes.

\subsection{On the importance of surface tension}\label{sec_st}

In the previous section we simplified our analysis in two ways, by assuming the system to be stationary in an inertial system where the line charge is at rest, and by neglecting surface tension. We will discuss semi-quantitatively the ramifications of this assumptions, with references to the literature where a further analysis of the full case is found. A full analysis of our present set-up with all effects included would be of considerable interest, but is beyond the scope of the present effort, although we intend to return to it in a future publication.

There exist several sources wherein various analyses of capillary-gravity waves generated by a moving disturbance may be found; embodiments include a general external pressure source or as a partially or totally immersed object \cite{raphael96, lamb75,chepelianskii08,chepelianskii10}. For comparison with literature, let us here make the common assumption $\rho_1=0, \rho_2=\rho$.

For capillary gravity waves on deep water one has in this case\cite{landau87}
\begin{equation}
\omega^2=gk+\sigma k^3/\rho. \label{A}
\end{equation}
The phase velocity $c=\sqrt{g/k+\sigma k/\rho}$ has a minimum when $k=\kappa$, where the wave number $\kappa=\sqrt{\rho g/\sigma}$ is the inverse capillary length, is
\begin{equation}
c_{\rm min}=\sqrt{\frac{2g}{\kappa}}=\left(\frac{4g\sigma}{\rho}\right)^{1/4}. \label{B}
\end{equation}
 When $k=\kappa$ the group velocity $c_g=d\omega/dk$ is the same as the phase velocity, $c_g=c=c_{\rm min}$. For $k \gg \kappa$ (ripples), the effect of gravity is negligible, and the group velocity becomes actually greater than the phase velocity, $c_g \rightarrow 3c/2$. When $\sigma\to 0$, however, $\kappa\to \infty$, and this regime is not available.

 Consider now the general situation above, where the pressure source in the form of a charged line moves at constant velocity $V$, above the surface and parallel to it. An important role is here played by the velocity $c_{\rm min}$: in the integral solutions, singular pole contributions occur when
 \begin{equation}
 \omega^2-V^2 k^2=0 \label{C}
 \end{equation}
 (cf., for instance,  Ref.~\cite{raphael96}). This can only happen only when $V>c_{\rm min}$. There are then two positive solutions to this equation, corresponding to gravity waves and capillary waves, respectively. The smaller wave number $k<\kappa$ (long wave) corresponds to a group velocity being less than $V$, so that this wave is formed behind the moving disturbance. The larger wave number $k>\kappa$ (short wave) corresponds to a group velocity greater than $V$, so that this wave is found in front of the moving disturbance. An important point is that for $V>c_{\rm min}$ the disturbance leads to a {\it wave resistance}, and the effect has a similarity to the \v{C}erenkov effect experienced by a charged particle moving superluminally in matter.  
 
 Assuming as we have that $c=V$ and $\sigma=0$ then leads to $c=V=\sqrt{g/k}$, which is exactly equation (\ref{lambda}), since $k=2\pi/\lambda$. It also means waves can have but a single frequency $\omega=g/V$. 
 
This answers a couple of questions that arose in the previous analysis: 
(1) Are there waves in front of the line charge? In the quasistationary inviscid case considered, group velocity does not appear play a role, since the wave group extends to infinity. However, the question of whether there will be a forward-propagating train of waves depends on whether there exists a physically acceptable solution with group velocity $>V$. We have seen that this is in principle always so when $V>c_\text{min}$, which is zero when $\sigma=0$.  To wit the two wave vector solutions become, in the limit of small $\sigma$ \cite{raphael96}
\be
  k \buildrel{\sigma\to 0}\over{\longrightarrow} \left\{\begin{array}{c}g/V^2 \\ \rho V^2/\sigma\end{array} \right..
\ee
The second of these (capillary waves) tends to infinity and can be ignored (one should be careful, since setting $\sigma=0$ is then not the same as taking the \emph{limit} $\sigma\to 0$). Mathematically the loss of the forward propagating solution is linked to the fact that setting $\sigma=0$ reduces the governing equation (\ref{bigEq}) from order 4 to order 2, having fewer solutions. (2) An infinite train of waves has infinite energy. However, when $V>c_\text{min}$ the moving charge feels a wave resistance, and a constant energy input is required to maintain constant speed, and stationary conditions require that the charged has been in constant motion for an eternity.

\section{Conclusions}

We have considered the basic electrohydrodynamic case of a point charge and a line charge above a fluid-fluid interface. The electric field from the charge acts on the dielectric interface, causing it to rise above the equilibrium level. In the static case the shape of the fluid-fluid interface is determined by the relative strength of three forces acting on it: the electromagnetic lifting force, gravity and surface tension.

For small elevations the deformation can be described by a single ordinary differential equation which is readily solved numerically. We derive and solve this equation for the case of a point charge and a line charge. When surface tension is small the maximum height of the surface is determined by a dimensionless number which is the relative weight of the electric lifting force vs.\ gravity. The effect of surface tension is to counteract the tendency of the other forces to induce surface curvature, which we illustrate in figure \ref{fig:surfacetension}.

Finally we derived the dynamical equation of motion for a surface subject to a line charge moving at a constant velocity. Altough apparently just a small generalization of the static case, the system is surprisingly rich in physics. As the moving line charge goes past, the liquid surfaces is elevated from its unperturbed level and is left to oscillate around an equilibrium level as the charge moves away [we assume inviscid fluid, so the oscillations are undampled]. The details of the time dependent movement of the fluid surface, however, depend nontrivially on the speed at which the charge moves, compared to a reference velocity $\sqrt{gL}$ [$L$ being the smallest distance from charge to undisturbed surface]. For high velocities the surface hardly has time to react to the presence of the charge, and the oscillations have long wavelength and small amplitude. For low velocities the oscillation wavelength is short, and the maximum elevation can be considerably higher than that of static equilibrium. The amplitude of oscillations behind the charge are small also in this extreme, however, since the charge has time to set the surface gently back behind it. In between these extremes is found the velocity giving the highest amplitude of oscillations behind the moving charge. 
We also discussed how surface tension influences the wave pattern. In particular, waves of group velocity greater than the phase velocity is possible, allowing waves to appear in front of the moving line charge.

\end{document}